\documentclass[10pt,journal,letterpaper,compsoc]{IEEEtran}
\ifCLASSOPTIONcompsoc
  \usepackage[nocompress]{cite}
\else
  \usepackage{cite}
\fi

\usepackage{amsfonts}
\usepackage{amsmath}
\ifCLASSINFOpdf
\else
\fi
\usepackage{graphicx}
\begin{document}
%
% paper title
% Titles are generally capitalized except for words such as a, an, and, as,
% at, but, by, for, in, nor, of, on, or, the, to and up, which are usually
% not capitalized unless they are the first or last word of the title.
% Linebreaks \\ can be used within to get better formatting as desired.
% Do not put math or special symbols in the title.
\title{Visual Analytics of Group Differences in Tensor Fields:
Application to Clinical DTI}
%
%
% author names and IEEE memberships
% note positions of commas and nonbreaking spaces ( ~ ) LaTeX will not break
% a structure at a ~ so this keeps an author's name from being broken across
% two lines.
% use \thanks{} to gain access to the first footnote area
% a separate \thanks must be used for each paragraph as LaTeX2e's \thanks
% was not built to handle multiple paragraphs
%
%
%\IEEEcompsocitemizethanks is a special \thanks that produces the bulleted
% lists the Computer Society journals use for "first footnote" author
% affiliations. Use \IEEEcompsocthanksitem which works much like \item
% for each affiliation group. When not in compsoc mode,
% \IEEEcompsocitemizethanks becomes like \thanks and
% \IEEEcompsocthanksitem becomes a line break with idention. This
% facilitates dual compilation, although admittedly the differences in the
% desired content of \author between the different types of papers makes a
% one-size-fits-all approach a daunting prospect. For instance, compsoc 
% journal papers have the author affiliations above the "Manuscript
% received ..."  text while in non-compsoc journals this is reversed. Sigh.

% ORCIDs:
% Amin:
% Vitalis: orcid.org/0000-0002-1270-8861
% Tobias: orcid.org/0000-0002-7929-1079
% Pia: orcid.org/0000-0001-9237-1236
% Reinhard: orcid.org/0000-0002-5505-9347 
% Thomas: orcid.org/0000-0002-1200-7248
\author{Amin Abbasloo, Vitalis Wiens, Tobias Schmidt-Wilcke, Pia Sundgren, Reinhard Klein, and Thomas Schultz% <-this % stops a space
\IEEEcompsocitemizethanks{\IEEEcompsocthanksitem A.~Abbasloo, V.~Wiens, and R.~Klein are with the Department of Computer Science, University of Bonn, Bonn, Germany.\protect\\
% note need leading \protect in front of \\ to get a newline within \thanks as
% \\ is fragile and will error, could use \hfil\break instead.
E-mail: \{abbasloo,wiens,rk\}@cs.uni-bonn.de
\IEEEcompsocthanksitem T.~Schmidt-Wilcke is with St.~Mauritius Hospital, Meerbusch and Institute of Clinical Neuroscience and Medical Psychology, University of D\"usseldorf, D\"usseldorf, Germany.\protect\\
E-mail: tobias-schmidt-wilcke@t-online.de
\IEEEcompsocthanksitem P.~Sundgren is with the Department of Radiology, University of Michigan Health Care, Ann Arbor, MI, USA and with the Department of Clinical Sciences/Radiology, Lund University, Lund, Sweden.\protect\\
E-mail: pia.sundgren@med.lu.se
\IEEEcompsocthanksitem T.~Schultz is with the Bonn-Aachen International Center for Information Technology (B-IT) and the Department of Computer Science, University of Bonn, Bonn, Germany.\protect\\
E-mail: schultz@cs.uni-bonn.de
}% <-this % stops a space
%\thanks{Manuscript received November ?, 2017.}
}

% note the % following the last \IEEEmembership and also \thanks - 
% these prevent an unwanted space from occurring between the last author name
% and the end of the author line. i.e., if you had this:
% 
% \author{....lastname \thanks{...} \thanks{...} }
%                     ^------------^------------^----Do not want these spaces!
%
% a space would be appended to the last name and could cause every name on that
% line to be shifted left slightly. This is one of those "LaTeX things". For
% instance, "\textbf{A} \textbf{B}" will typeset as "A B" not "AB". To get
% "AB" then you have to do: "\textbf{A}\textbf{B}"
% \thanks is no different in this regard, so shield the last } of each \thanks
% that ends a line with a % and do not let a space in before the next \thanks.
% Spaces after \IEEEmembership other than the last one are OK (and needed) as
% you are supposed to have spaces between the names. For what it is worth,
% this is a minor point as most people would not even notice if the said evil
% space somehow managed to creep in.

% The paper headers
\markboth{Submitted to IEEE Transactions on Visualization and Computer Graphics}%
{Abbasloo \MakeLowercase{\textit{et al.}}: Integrated Statistical and Visual Analysis of Group Differences in Tensor Fields}
% The only time the second header will appear is for the odd numbered pages
% after the title page when using the twoside option.
% 
% *** Note that you probably will NOT want to include the author's ***
% *** name in the headers of peer review papers.                   ***
% You can use \ifCLASSOPTIONpeerreview for conditional compilation here if
% you desire.

% The publisher's ID mark at the bottom of the page is less important with
% Computer Society journal papers as those publications place the marks
% outside of the main text columns and, therefore, unlike regular IEEE
% journals, the available text space is not reduced by their presence.
% If you want to put a publisher's ID mark on the page you can do it like
% this:
%\IEEEpubid{0000--0000/00\$00.00~\copyright~2015 IEEE}
% or like this to get the Computer Society new two part style.
%\IEEEpubid{\makebox[\columnwidth]{\hfill 0000--0000/00/\$00.00~\copyright~2015 IEEE}%
%\hspace{\columnsep}\makebox[\columnwidth]{Published by the IEEE Computer Society\hfill}}
% Remember, if you use this you must call \IEEEpubidadjcol in the second
% column for its text to clear the IEEEpubid mark (Computer Society journal
% papers don't need this extra clearance.)

% use for special paper notices
%\IEEEspecialpapernotice{(Invited Paper)}

% for Computer Society papers, we must declare the abstract and index terms
% PRIOR to the title within the \IEEEtitleabstractindextext IEEEtran
% command as these need to go into the title area created by \maketitle.
% As a general rule, do not put math, special symbols or citations
% in the abstract or keywords.
\IEEEcompsoctitleabstractindextext{%
\begin{abstract}
We present a visual analytics system for exploring group differences in tensor fields with respect to all six degrees of freedom that are inherent in symmetric second-order tensors. Our framework closely integrates quantitative analysis, based on multivariate hypothesis testing and spatial cluster enhancement, with suitable visualization tools that facilitate interpretation of results, and forming of new hypotheses. Carefully chosen and linked spatial and abstract views show clusters of strong differences, and allow the analyst to relate them to the affected structures, to reveal the exact nature of the differences, and to investigate potential correlations. A mechanism for visually comparing the results of different tests or levels of smoothing is also provided.

We carefully justify the need for such a visual analytics tool from a practical and theoretical point of view. In close collaboration with our clinical co-authors, we apply it to the results of a diffusion tensor imaging study of systemic lupus erythematosus, in which it revealed previously unknown group differences.
\end{abstract}

% Note that keywords are not normally used for peerreview papers.
}

% make the title area
\maketitle

% To allow for easy dual compilation without having to reenter the
% abstract/keywords data, the \IEEEtitleabstractindextext text will
% not be used in maketitle, but will appear (i.e., to be "transported")
% here as \IEEEdisplaynontitleabstractindextext when compsoc mode
% is not selected <OR> if conference mode is selected - because compsoc
% conference papers position the abstract like regular (non-compsoc)
% papers do!
\IEEEdisplaynontitleabstractindextext
% \IEEEdisplaynontitleabstractindextext has no effect when using
% compsoc under a non-conference mode.

% For peer review papers, you can put extra information on the cover
% page as needed:
% \ifCLASSOPTIONpeerreview
% \begin{center} \bfseries EDICS Category: 3-BBND \end{center}
% \fi
%
% For peerreview papers, this IEEEtran command inserts a page break and
% creates the second title. It will be ignored for other modes.
\IEEEpeerreviewmaketitle

\ifCLASSOPTIONcompsoc
  \noindent\raisebox{2\baselineskip}[0pt][0pt]%
  {\parbox{\columnwidth}{\section{Introduction}\label{sec:introduction}%
  \global\everypar=\everypar}}%
  \vspace{-1\baselineskip}\vspace{-\parskip}\par
\else
  \section{Introduction}\label{sec:introduction}\par
\fi
\IEEEPARstart{D}{iffusion} Tensor Imaging estimates a symmetric second-order tensor field that encodes the direction and strength of water molecule diffusion inside biological tissue. It is widely used for clinical studies of brain disease, because it makes it possible to measure parameters that relate to tissue microstructure and that correlate with factors such as the integrity of neural tracts \cite{Basser:1996}.

In this context, one of the most widely used methods for data analysis
are statistical hypothesis tests that aim to show significant differences between specific populations, such as comparing patients suffering from some type of disease to healthy controls \cite{Cercignani:2010,o2015statistical}. In current practice, univariate tests, which only account for a single parameter at each point of the tensor field, such as Fractional Anisotropy or Mean Diffusivity, are by far the most widely used.

It has been argued by many that reducing the tensor field to just one or two scalar fields eliminates much of the available information, and could thus cause the analysis to miss some of the group differences. In order to simultaneously account for all degrees of freedom, multivariate statistical methods have been developed based on a tensor normal distribution~\cite{basser2003normal}, on interpoint distances~\cite{baringhaus2004new, whitcher2007statistical}, or on eigenvalues and eigenvectors~\cite{schwartzman2010group}.

\begin{figure*}
  \centering
  \includegraphics[width=\textwidth]{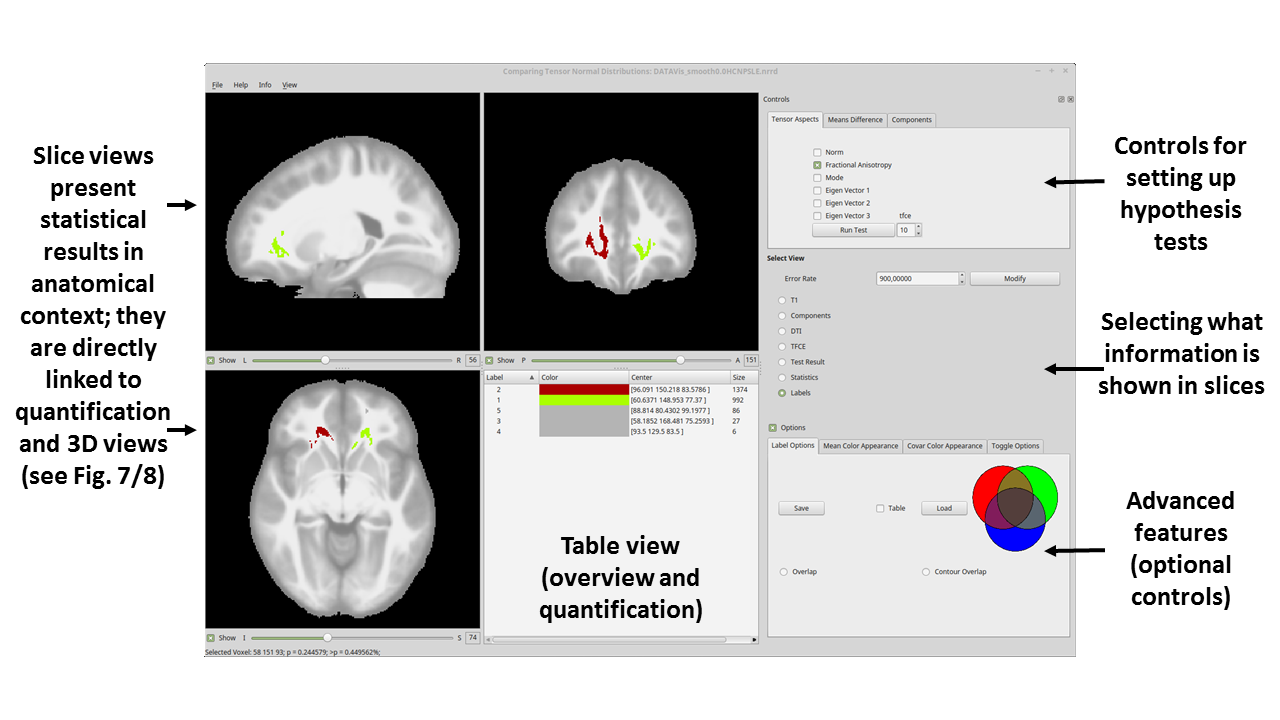}
  \caption{The main window of our visual analytics system includes
    slice views, a table view (here, used to highlight two specific
    regions with significant group differences in red and light green), and
    controls that allow the user to perform different hypothesis
    tests, as well as to explore and compare their results with a
    range of visualization tools.}
  \label{fig:teaser}
\end{figure*}

In Section~\ref{sec:motivation}, we argue that it is more suitable to consider the detection of group differences in tensor fields as a visual analytics problem, rather than as a purely statistical task. To our knowledge, we present the first visual analytics pipeline that addresses two important reasons why the potentially more powerful multivariate tests have so far not been widely adopted in practice:
\begin{enumerate}
\item Even though one might hope that testing a more general null hypothesis, e.g., testing for arbitrary differences in the tensor fields, will increase the number and spatial extent of differences one is able to detect, the opposite can happen when dealing with limited sample sizes, as it is usually the case in clinical studies. This effect will be illustrated in Section~\ref{sec:motivation}. The reason for it is that adding degrees of freedom that contain variability, but no strong group difference, will reduce rather than increase the power of a statistical test.
\item Most multivariate tests point out locations at which groups of tensor fields differ, but do not provide any detail on which tensor attributes are affected, which makes their results difficult to interpret. When testing more specific properties, such as Fractional Anisotropy, more specific conclusions can be drawn in terms of the corresponding differences in the biological tissue.
\end{enumerate}

To address the first point, our system, which can be seen in Figure~\ref{fig:teaser}, allows the analyst to interactively build and test a wide range of interpretable null hypotheses that can span the full spectrum between focusing on individual degrees of freedom to including all available information. In an exploratory process, the analyst can discover various types of group differences, and identify a null hypothesis that appears to be most sensitive for the specific type of disease. In order to maintain full statistical validity, a small number of final hypotheses should result from this exploratory analysis, and should later be tested on independent data.

To address the second point, our system provides a mechanism to investigate which specific attributes of the tensor fields have caused a multivariate test to indicate a group difference. In contrast to the \emph{a posteriori} characterization proposed quite recently~\cite{bouchon2016most}, our method accounts for all six degrees of freedom, rather than just axial diffusion, radial diffusion, and Fractional Anisotropy, and it provides visual support for studying correlations between them.

After describing our system in Section~\ref{sec:framework}, we report novel insights that it enabled on data from a clinical study of systemic lupus erythematosus in Section~\ref{sec:results}, and conclude the paper with a discussion in Section~\ref{sec:conclusion}.

%%% Local Variables:
%%% mode: latex
%%% TeX-master: "../submission"
%%% End:

\section{Related Work}

Our work can be considered as an example of comparative visualization
\cite{Pagendarm:1995}. In this respect, the most closely related works
are the Tender glyph, which explicitly encodes differences between two
symmetric positive definite tensors \cite{zhang2016glyph}, and recent
followup works that have extended this encoding and combined it with
complementary visualizations to enable visualization of tensor field
ensembles \cite{zhang:2017ensemble}, as well as a visual comparison of
groups of tensor fields at multiple levels of detail
\cite{Zhang:2017group}. Unlike this existing approach, which focuses
on a visual assessment, our goal is to closely integrate visual with
quantitative statistical analysis. This represents a significant
extension of our own previous work on visualizing tensor normal
distributions \cite{abbasloo2016visualizing}, which did not include
any mechanisms for group comparison. Our integrated approach will be
motivated in more detail in Section~\ref{sec:motivation} and leads to
new and different design choices.

Seen in a wider context, our work contributes to addressing an urgent
need for visual analytics technology that helps us to handle the
increasing volume, variety, and velocity of image data that is used in
medical and biological studies. Other recent examples include a series
of works for interactive visualization and segmentation proofreading
in connectomics research \cite{Pfister2014}, tools that support
navigation of parameter spaces for automated image analysis
\cite{Pretorius:2011,TorsneyWeir:2011,Schultz:VAST2013}, a tool by
Klemm et al., which supports hypothesis generation in epidemiology by
integrating image-based with non-image data \cite{Klemm2014}, and one by
Hermann et al., which allows biologists to explore differences in
anatomical shapes based on interactively defined groups of 3D images
\cite{Hermann2016}.%, and one by Gorgi Zadeh et al., who provide a tool
%to gain a quick overview of a large number of images from
%high-throughput microscopy with respect to the sizes and densities of
%depicted clusters \cite{Zadeh2016}.

Our proposed tool makes use of several state-of-the-art techniques for
diffusion tensor analysis, including for spatial normalization
\cite{zhang2007high,zhang2007unbiased}, interpretable
reparametrization \cite{kindlmann2007diffusion}, and cluster
enhancement of the resulting statistical maps
\cite{smith2009threshold}. Details on these will be provided as needed
throughout the paper.

%%% Local Variables:
%%% mode: latex
%%% TeX-master: "../submission"
%%% End:

\begin{figure*}
 \centering
\includegraphics[scale=0.5]{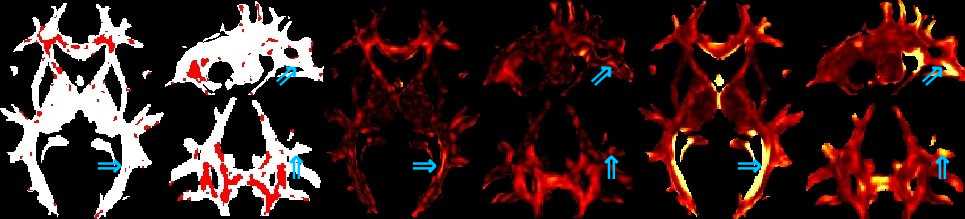}
 \caption{A group difference at a particular location is considered statistically significant (marked red in left subfigure) if the magnitude of the group difference (color coded in the center), is large relative to the local variance (color coded on the right). Blue arrows highlight some regions in which a strong difference fails to be significant due to large variance.}
 \label{fig:cfaall}
\end{figure*}

\section{Finding Group Differences in Tensor Fields as a Visual Analytics Problem}
\label{sec:motivation}

Our current work arose out of a previous collaboration between some of the computer science and clinical co-authors, which resulted in a new statistical and machine learning technique \cite{Khatami:2017}, and in findings that have been reported in clinical journals, e.g.,~\cite{schmidt2014diminished}. In-depth discussions about the statistical analysis of tensor fields led us to conclude that there is a need for a visual analytics approach that tightly integrates formal quantitative with visual analysis. We will provide some theoretical background and intuition in Section~\ref{sec:background}, followed by a requirements analysis in Section~\ref{sec:task-analysis}.

\subsection{Background on Multivariate Hypothesis Testing}
\label{sec:background}

A direct visual encoding that supports the comparison of groups of tensor fields has been proposed very recently \cite{Zhang:2017group}. When applied to clinical data, such a strategy will show at least small differences everywhere: Due to noise and other random effects, even two very similar groups of healthy adults are very unlikely to be exactly the same. A benefit of integrating visual and statistical analysis, which is an important goal of our system, is that it allows us to focus on differences that are likely to be due to the studied disease.

Intuitively, statistical hypothesis tests decide on the significance of specific differences by comparing their magnitude to the amount of variance in the data. This is illustrated in Figure~\ref{fig:cfaall}: The red regions in the left part, in which statistically significant differences aligned with FA have been detected, have a clear correlation with regions of strong group difference, which is color coded in the central part. However, there are regions with a strong group difference, some of them highlighted with blue arrows, that are not significant in a statistical sense. This happens where the variance, color coded in the right part of the figure, is so large that even large group differences might have arisen due to chance.

\begin{figure}
  \centering
  \begin{minipage}[t]{0.47\linewidth}
  \includegraphics[width=\linewidth]{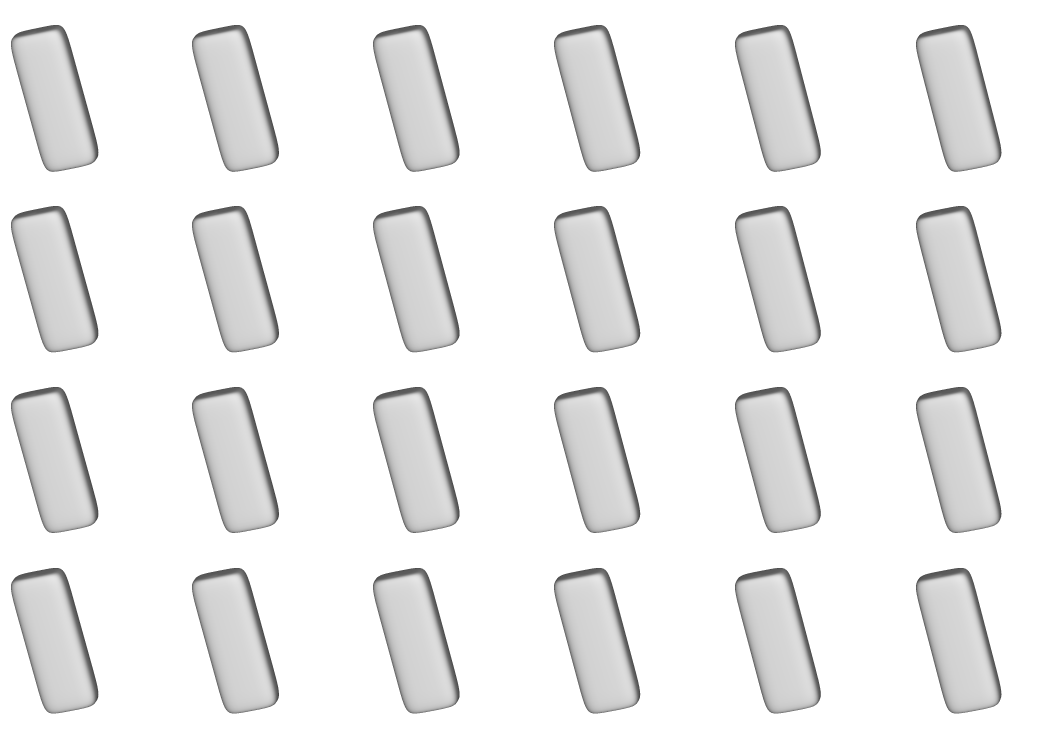}
  \centering\footnotesize Synthetic Tensor Field A
  \end{minipage}\hspace{1ex}
  \hskip0.03\linewidth
  \begin{minipage}[t]{0.47\linewidth}
  \includegraphics[width=\linewidth]{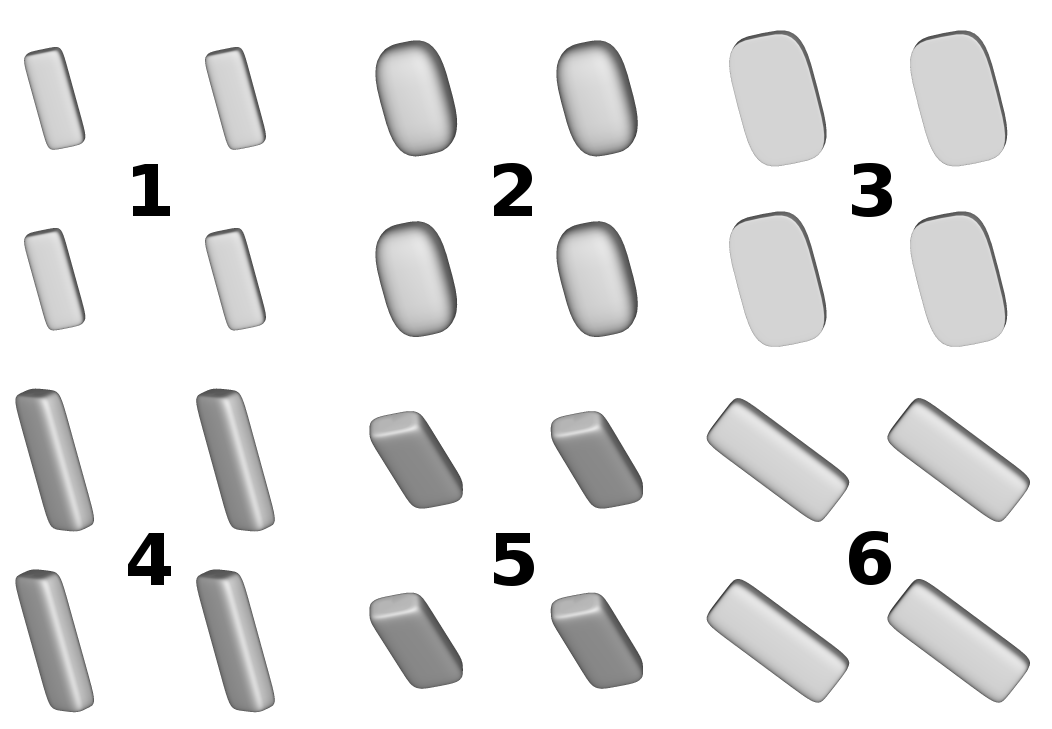}
  \centering  \footnotesize Synthetic Tensor Field B
  \end{minipage}
  \centering
  \caption{The two tensor fields A and B have been designed to differ
    at all locations. However, a commonly used statistical hypothesis test based
    on FA alone would only detect the difference in region 2. We
    provide a visual analytics tool to explore all types of
    differences.}
  \label{fig:synthetic-tensor-fields}
\end{figure}

Most clinical studies perform hypothesis testing on only one or two
scalar values derived from the tensor field, such as Fractional
Anisotropy (FA) or Mean Diffusivity (MD). To illustrate the need for
multivariate hypothesis tests,
Figure~\ref{fig:synthetic-tensor-fields} presents two synthetic tensor
fields. It is obvious from their visualization using superquadric
glyphs \cite{Kindlmann:2004} that the two fields differ at all
locations, with respect to all six degrees of freedom in $3\times 3$
symmetric tensors: Reduced overall magnitude, as measured by reduced
Frobenius norm, i.e., $\ell_2$ norm of eigenvalues (1); reduced degree of
directional dependence, as measured by reduced FA
(2); change from a more linear towards a more planar type of
anisotropy as measured by reduced tensor mode, while keeping FA
constant (3), as well as rotations around three orthogonal axes
(4--6).

In this example, the widely used univariate hypothesis test based on FA would be
blind to all differences except those in region 2. In contrast to this, a multivariate test that accounts for all available information in the tensor should be able to detect all different types of variation. Accordingly, multivariate testing is often applied in the hope that it will be able to detect more and larger regions of significant differences. For example, as a key result of their proposed multivariate Cram{\'e}r test, Whitcher et al.\ report that they observed a 169\% increase ``in the volume of a significant cluster compared to the univariate FA test'' \cite{whitcher2007statistical}.

\begin{figure}
 \centering
\includegraphics[scale=0.22]{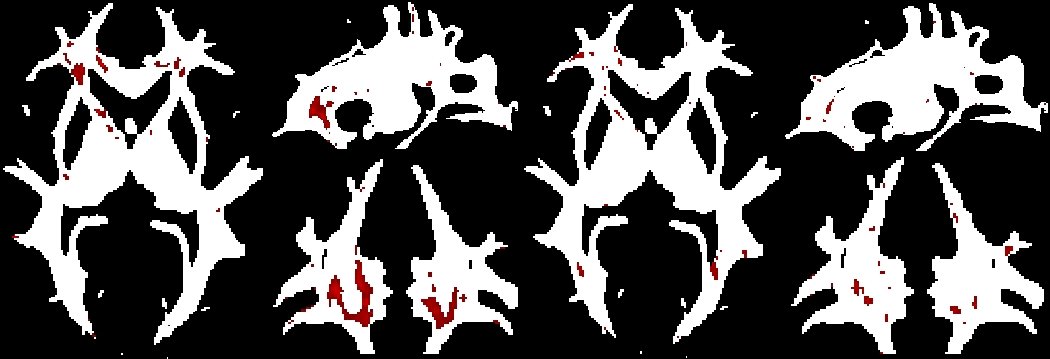}
\caption{Compared to a univariate test of tensor changes associated
  with FA (left), a multivariate test for arbitrary changes in the
  tensor (right) produces fewer, rather than more, significant
  results.}
 \label{fig:faall}
\end{figure}

Disappointingly, in our own experiments on clinical data, we often
observed a decrease, rather than an increase of the overall number of
significant voxels when replacing the widely used univariate $t$-tests
with its multivariate counterpart, Hotelling's $T^2$ test. An example
is shown in Figure~\ref{fig:faall}. The tests will be explained in
more detail in Section~\ref{sec:igrt-projections}. For now, we observe
on the left that clear and extended clusters (red) have been detected
when looking for changes along one specific axis in tensor space,
namely changes in Fractional Anisotropy (FA). One might hope that a
test that accounts for the full tensor information would highlight the
same regions, plus others. Unfortunately, it is clear from the result
on the right that this is not the case. We note that even the 169\%
increase in volume reported in \cite{whitcher2007statistical} pertains
to \emph{a single selected cluster;} the authors do not report whether, and
by how much, the overall volume of all clusters in their data
increased.

\begin{figure}[t]
  \centering
  \begin{minipage}[t]{0.48\linewidth}
  \includegraphics[width=\linewidth]{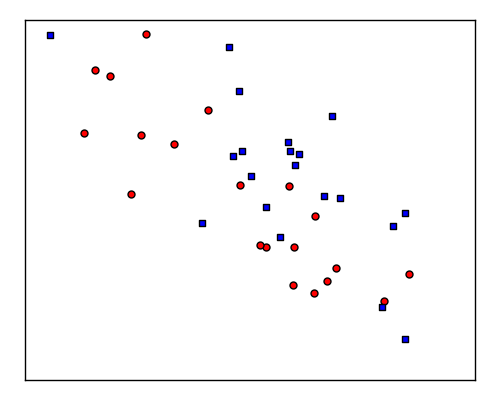}
  \centering\footnotesize (a)
  \end{minipage}\hspace{1ex}
  \begin{minipage}[t]{0.48\linewidth}
  \includegraphics[width=\linewidth]{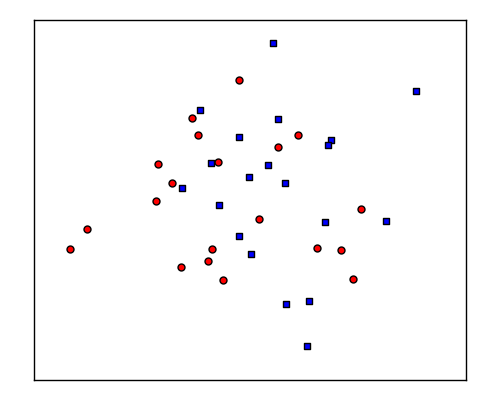}
  \centering  \footnotesize (b)
  \end{minipage}
  \centering
  \caption{In~(a), a multivariate Hotelling test that accounts for
    both axes is more powerful for detecting the difference between
    red circles and blue squares than univariate $t$ tests along
    either axis. In~(b), a univariate test along the horizontal axis
    is more powerful than a multivariate test. Axes are not labeled,
    since the effect is unaffected by shifting and uniform scaling.}
  \label{fig:illustrative-scatter-plots}
\end{figure}

To better understand why, depending on the data characteristics, a
multivariate test may or may not be more powerful than a simple
univariate test, Figure~\ref{fig:illustrative-scatter-plots} presents
two toy examples. In the first one, the blue squares are sampled from
a multivariate Gaussian distribution whose center is at the top right
compared to the center of the distribution from which the red circles
are taken. Variance along both axes is so high that, with a univariate
$t$ test, neither horizontal nor vertical distance alone is sufficient
to detect a significant difference ($p>0.11$ and $p>0.43$,
respectively). In contrast, by accounting for the differences in both
dimensions simultaneously, a Hotelling test achieves a clearly
significant result ($p<0.004$).

In the second example, the center of the blue squares is to the right
compared to the red circles, but at the same height. Here, a $t$ test
along the horizontal axis shows a significant difference ($p<0.03$),
but the Hotelling test fails to produce a significant result
($p>0.06$). The reason is that, in this case, including the second
dimension adds pure noise, which reduces the power of the test. Of
course, if we increase the number of samples sufficiently, also the
multivariate test will indicate a significant difference. However, due
to the fact that many clinical studies face difficulties in finding a
large number of patients that can be included, and due to the high
cost of brain imaging, we have to carefully pick our null hypothesis
in a way that the resulting test will be sensitive enough even for a
very limited number of samples.

% Consequently, even Bouchon et al., who
%advocate the use of multivariate tests on diffusion tensors
%\cite{bouchon2016most}, conclude that ``if one has a strong \emph{a
%  priori} about the kind of expected change, it is preferable to
%resort to the scalar-based approach on the corresponding scalar
%image.''

When comparing tensor fields, the number of possible null hypotheses is immense, especially if we consider not just individual tensor properties, but also null hypotheses that combine several degrees of freedom, such as ``The two groups do not differ with respect to the amount or type of anisotropy'' (which requires a 2D multivariate test) or ``The two groups do not differ with respect to orientation'' (which requires a multivariate test in 3D orientation space). Since the choice of sensible null hypotheses should be guided both by the data and by the domain knowledge of the medical expert, we consider visual analytics as the only viable approach.

\subsection{Requirements Analysis}
\label{sec:task-analysis}

Based on our initial experiments with multivariate hypothesis testing and discussions with our clinical co-authors, we identified the following six requirements that our visual analytics system should meet:

\begin{itemize}
\item[\textbf{R1}] The system should implement state-of-the-art techniques for statistical hypothesis testing. It should make it easy to steer them, and to explore and interpret their results. The need for a close integration of statistical methods results from the fact that publishing findings from group studies in clinical journals requires a formal and quantitative data analysis.
\item[\textbf{R2}] The system should offer the two-dimensional slice views that clinical researchers are used to and that are frequently found in clinical publications.
\item[\textbf{R3}] The system should also provide three-dimensional views when it helps assess three-dimensional anatomical structures, e.g., via fiber tracking.
\item[\textbf{R4}] To facilitate interpretation of the results from multivariate hypothesis tests, the system should include visualizations that reveal which tensor properties differed most substantially between groups, and whether different properties were correlated.
\item[\textbf{R5}] In case of findings that might be surprising and difficult to explain, the system should support investigating the potential role of residual misalignment from subject normalization.
\item[\textbf{R6}] The system should support the direct comparison of the results from different null hypotheses, to judge the extent to which the corresponding regions might spatially overlap.
\end{itemize}

%%% Local Variables:
%%% mode: latex
%%% TeX-master: "../submission"
%%% End:

\section{Visual Analytics of Tensor Field Group Differences: A
  Practical Framework}
\label{sec:framework}

\subsection{Preprocessing}
\label{sec:preprocessing}

In order to compare all tensor fields on a per-voxel basis, we first
have to bring them into spatial correspondence by nonlinear
registration to a common template. As part of this process, tensors
were rotated according to the incurred changes of their frame of
reference. We used the publicly available Diffusion Tensor Imaging
ToolKit (DTI-TK) \cite{zhang2007high,zhang2007unbiased} to achieve
this.

Our visual analytics framework was implemented in C++ with Qt for the
user interface, Teem (teem.sf.net) for standard tensor visualization
including fiber tracking, and OpenGL for 3D graphics. Data
preprocessing, such as transforming all tensors into the IGRT basis
(Section~\ref{sec:igrt-projections}), or computation of means and
covariances of each subgroup, was performed using in-house Python
scripts.

\subsection{Steerable Statistical Hypothesis Testing}
\label{sec:hypothesis-testing}

As an important aspect of requirement \textbf{R1,} the user should be able to steer statistical hypothesis tests on tensor fields in a meaningful way. We achieve this by combining the multivariate Hotelling test with a specific reparameterization of the tensor field \cite{kindlmann2007diffusion}. Both techniques have been proposed previously, but combining them to achieve steerable hypothesis testing is one of our contributions.

\subsubsection{Multivariate Testing with Meaningful Projections}
\label{sec:igrt-projections}

The most common way to identify group differences in imaging studies is \emph{mass univariate testing.} This amounts to running a large number of statistical tests, each accounting only for a single value at a single location of the brain.  Spatially mapping locations where the null hypothesis was rejected highlights regions in which the groups differ in a statistically significant manner \cite{o2015statistical}.

Tensors are intrinsically multivariate, reflecting not just a single property (such as amount of anisotropy), but also the overall amount of diffusion, type of anisotropy, and preferred diffusion directions. Our framework accounts for this by testing more complex null hypotheses, stating that the mean tensors in the two groups are the same with respect to multiple or even all their properties. The Hotelling test provides the corresponding extension of the widely used $t$ test \cite{Srivastava:2002}. Its $T^2$ statistics is defined as
\begin{equation}
\label{eq:t2-statistic}
T^2=\frac{n_1n_2}{n_1+n_2}( \bar{\mathbf{d}}_1-\bar{\mathbf{d}}_2)^T \mathbf{S}_p^{-1} (\bar{\mathbf{d}}_1-\bar{\mathbf{d}}_2)
\end{equation} 
where $\bar{\mathbf{d}}_1$, $\bar{\mathbf{d}}_2$ are group mean vectors and $\mathbf{S}_p$ is a pooled estimate of the covariance matrix, $\mathbf{S}_p = \frac{(n_1-1)\mathbf{S}_1+(n_2-1)\mathbf{S}_2}{n_1+n_2-2}$, where $n_i$ and $\mathbf{S}_i$ are the number of subjects and sample covariance for group $i$, respectively. To apply this test to symmetric $3\times 3$ diffusion tensors with coefficients $D_{ij}$, we embed them isometrically into $\mathbb{R}^6$ by setting
\begin{equation}
  \label{eq:isometric-embedding}
  \mathbf{d}:=\left[D_{11}, D_{22}, D_{33}, \sqrt{2}D_{12}, \sqrt{2}D_{13}, \sqrt{2}D_{23}\right]^T.
\end{equation}

As we demonstrated in Section~\ref{sec:background}, blindly applying a multivariate test to all available degrees of freedom can lead to a loss of sensitivity due to a swamping with noise. To avoid this, our visual analytics framework allows the analyst to interactively formulate null hypotheses that relate to a meaningful subset of the information contained in the tensor fields, and that deliberately exclude other aspects. Unfortunately, the components in Eq.~(\ref{eq:isometric-embedding}) are not suitable for this task, since they depend on the chosen frame of reference, and lack an intuitive interpretation.

We address this by expressing tensors in the so-called IGRT (``invariant gradient and rotation tangent'') basis, a local orthonormal basis
that can be constructed from the gradients of tensor invariants such a tensor norm and Fractional Anisotropy, and from the tangent vectors corresponding to infinitesimal rotations \cite{kindlmann2007diffusion}. In particular, after rotating difference vectors $\mathbf{d}-\mathbf{\bar{d}}$ between individual tensor values $\mathbf{d}$ and the grand mean $\bar{\mathbf{d}}=\left(n_1\bar{\mathbf{d}}_1+n_2\bar{\mathbf{d}}_2\right)/(n_1+n_2)$ into the IGRT basis constructed at $\bar{\mathbf{d}}$, we can interpret their six rotated coefficients as changes in tensor value related to
\begin{itemize}%[leftmargin=20pt,itemsep=0.1ex]
\item[1.] Changes in Frobenius norm ($R_1=\|\mathbf{D}\|_F$)
\item[2.] Changes in the amount of anisotropy, as measured by $R_2=\mathrm{FA}$
\item[3.] Changes in the type of anisotropy, as measured by $R_3=\mathrm{mode}$
\item[4--6.] Rotations around the three eigenvectors
\end{itemize}

We allow the analyst to build interpretable null hypotheses by interactively selecting arbitrary combinations of these six degrees of freedom, which are illustrated in Figure~\ref{fig:synthetic-tensor-fields}.
Two properties of the IGRT basis are particularly relevant to its use in multivariate hypothesis testing, which goes beyond its previous use in visualization \cite{abbasloo2016visualizing}:

First, coefficients in the rotated difference vectors denote differences in tensor value (associated with changes in norm, FA, etc.), and thus all have the physical units of diffusivity. This means that they are measurements on a common scale, which is useful for combining them into a single multivariate test. However, it also implies a -- somewhat subtle -- mathematical difference between our framework, which tests FA-related tensor changes, and the more widely used practice of testing pre-computed FA values.

Second, Kindlmann et al.\ \cite{kindlmann2007diffusion} describe two variants of the IGRT basis: The one that was used in \cite{abbasloo2016visualizing} is derived from a cylindrical coordinate system, whose invariants are written as $K_i$ and include tensor trace, which is proportional to the widely used Mean Diffusivity (MD). In this work, we use the second one, which is derived from a spherical coordinate system, written as $R_i$, and includes the widely used Fractional Anisotropy (FA). Unfortunately, the gradients of MD and FA are non orthogonal, making it impossible to combine both into a common orthonormal IGRT frame. We have chosen the one that includes FA rather than MD to facilitate a direct comparison with a previous work that has tested FA on the same data \cite{schmidt2014diminished}.

\subsubsection{Cluster Enhancement}
\label{sec:tfce-enhancement}

Requirement \textbf{R1} states that our statistical tests should conform to the state of the art in neuroimage analysis. This requires addressing the problem of \emph{multiple comparisons.} When designing a statistical hypothesis test, we have to set its type~I error rate $\alpha$. In our context, a type~I error amounts to reporting a group difference in a region where there appears to be one in the sample -- e.g., due to noise or due to the specific choice of subjects -- even though there is none in the underlying populations. In neuroimaging, it is widely accepted to use $\alpha=0.05$.

Performing a statistical test in a huge number of voxels greatly increases the \emph{family-wise error,} i.e., the probability that the null hypothesis will be falsely rejected in at least one voxel. This could be addressed by reducing the $\alpha$ of each individual test, a strategy known as Bonferroni correction. However, this approach is overly conservative when nearby voxels are correlated, as it is typically the case in practice, and leads to an unnecessarily drastic reduction of $\alpha$ that no longer allows us to detect any of the true group differences.

A common way to correct for family-wise errors while preserving a useful amount of sensitivity is to rely on the size of contiguous regions in which an effect is observed, since large regions are less likely to occur due to chance \cite{nichols2003controlling}. Our framework implements threshold-free cluster enhancement (TFCE) \cite{smith2009threshold}, a state-of-the-art variant of this idea. TFCE automatically tries out a wide range of possible thresholds to form contiguous regions, then sums up the volumes of the clusters that each voxel belongs to at all these thresholds, so that voxels that belong to larger neighborhoods receive greater values without the need to select any specific cluster forming treshold.

Usually, TFCE would be used within the framework of a permutation based statistical test \cite{nichols2002nonparametric}, which involves computing the test statistic and its TFCE transformation for a large number of randomly permuted group labels. Based on this procedure, one can determine a value at which the TFCE map should be thresholded to yield a statistical test with a given family-wise error rate. Unfortunately, permutation based testing can take many hours on a standard workstation, and is thus much too time-consuming for use in an interactive visual analytics framework.

\begin{figure}
  \centering
  \includegraphics[width=\linewidth]{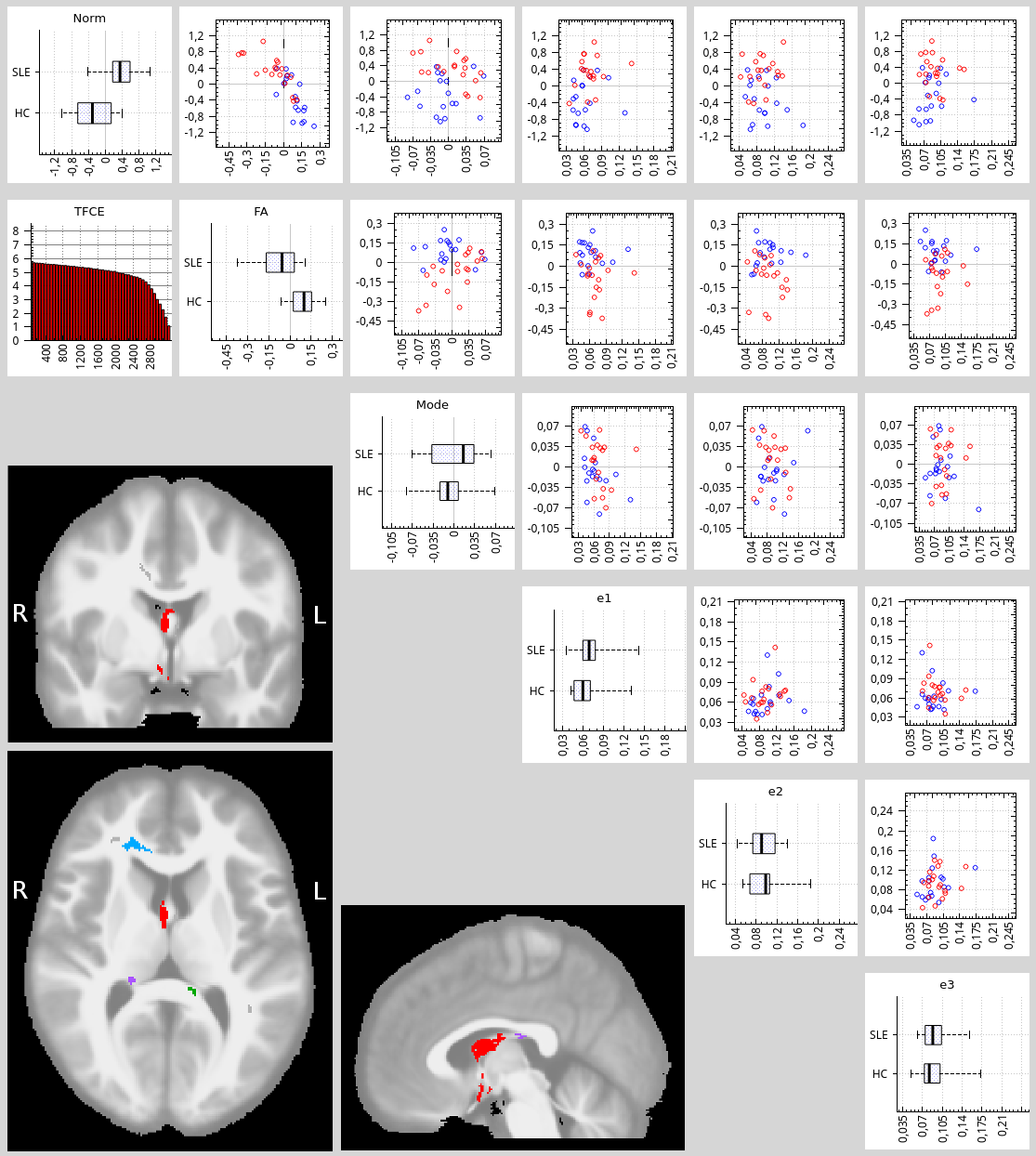}
  \caption{A scatter plot matrix allows the analyst to identify which tensor properties led to an overall difference being detected in a specific region (in this example, the one highlighted in red in the inset figures). It also shows correlations between different tensor attributes in the respective region. Tooltips complement the visual analysis with quantitative results.}
  \label{fig:splom}
\end{figure}

Within an exploratory analysis, which is the main task of our tool, it is anyway quite common to look at results not just for a single setting of $\alpha$, but to explore different ``levels of significance''. Therefore, we allow the analyst to threshold TFCE maps at arbitrary values, guided by a cumulative histogram that visualizes how the number of super-threshold voxels depends on the threshold value. Once the analyst has found a set of interesting clusters in an interactive session and would like to determine their exact family-wise error corrected $p$ values, he or she can run the permutation tests off-line. An important benefit of our framework is that it allows him or her to get a quick impression of the results for many potential hypotheses, and to easily discard those that are obviously fruitless.

\subsection{Spatial Views}
\label{sec:spatial-views}

After specifying and running a hypothesis test, the analyst would like to investigate the anatomical location, spatial extent, and shape of the regions in which a group difference was detected. According to requirement \textbf{R2,} axis-aligned slice views play a prominent role in our user interface and in many of our results. Our slices follow the radiological convention, i.e., the patient's right hemisphere is shown on the left hand side of each picture. Labels for left and right are included as a reminder of this (e.g., see Figure~\ref{fig:splom}).

\begin{figure}
 \centering
 \includegraphics[width=0.95\linewidth]{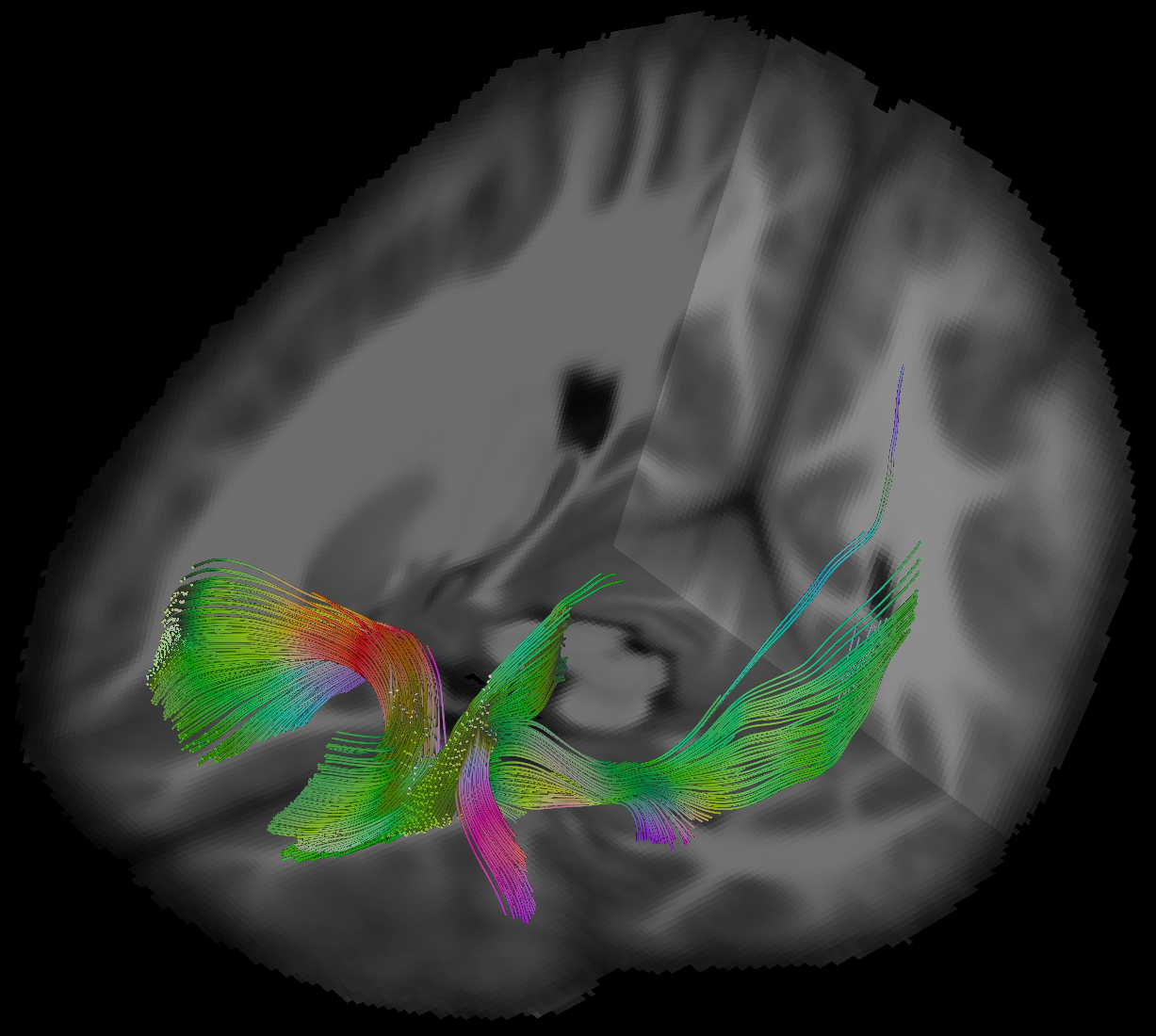}
 \caption{Seeding streamline tractography in a region of differences
   between the tensor fields allows the analyst to quickly identify the
   affected fiber tracts based on their characteristic
   shapes and anatomical context.}
 \label{fig:fibers}
\end{figure}

Statistical analysis involves a spatial warping of the involved subjects to a common reference space. By default, anatomical context for the detected regions is provided by superimposing them on averaged anatomical MR images that have been transformed into the same reference space. On demand, a more exact assessment of the affected tract
can be made by displaying an XYZ-RGB color encoding of the principal
diffusion direction \cite{Pajevic:1999}.

In agreement with requirement \textbf{R3,} our system also supports
seeding a streamline-based tractography algorithm \cite{Basser:2000a}
in an affected region, and provides a three-dimensional view of the
result, with orthogonal slices as optional anatomical context. In
order not to bias the tractography towards any of the involved
subjects, we run it on an average of all coregistered tensor
fields. The average is taken in Log-Euclidean space
\cite{Arsigny:2006}, which is known to minimize blurring
\cite{Keihaninejad2013}. We found that this strategy preserves the
characteristic shapes of the major bundles. For example, parts of the
forceps minor, anterior thalamic radiation, inferior fronto-occipital
fasciculus, and uncinate fasciculus are well-recognizable in
Figure~\ref{fig:fibers}.

\subsection{Table View}
\label{sec:table-views}

In neuroimaging, contiguous sets of voxels in which a significant difference was detected are referred to as clusters. Quickly and objectively assessing the number and volume of all detected clusters is part of requirement \textbf{R1,} and we support it by displaying this quantitative information in a table, sorted by cluster size. This table can be seen in the immediate neighborhood of the slice views in Figure~\ref{fig:teaser}. It also includes spatial information (center of gravity of each cluster) and is linked to the slices, so that clicking on a table row moves the slices to the respective cluster center.

To highlight and better distinguish clusters of particular interest in the spatial views, the user may assign a color to them in the table view. In this mode, unselected clusters are shown in gray, to make them less visually salient than the more relevant ones, while still distinguishing them from the rest of the brain and the background. An example can be seen in the inset pictures in Figure~\ref{fig:splom}.

\subsection{Scatter Plot Matrix}
\label{sec:splom}

The spatial views indicate where a user-defined null
hypothesis has been rejected, but not for which reason, which is the main concern of requirement \textbf{R4.} For example, when viewing the results of the null hypothesis ``The two groups do not differ with respect to any tensor property,'' the analyst will want to drill down into the
exact nature of the detected group differences, which requires
viewing all local properties of all involved subjects.

In tensor field visualization, glyphs are commonly used to show all available information at a given location \cite{Schultz:Vis10}, and our system offers a glyph view that will be discussed in the next subsection. Even though glyphs have been proposed that simultaneously visualize variations in tensor scale, shape, and orientation \cite{zhang:2017ensemble}, these glyphs are based on aggregating some of the available information. In particular, they do not encode correlations between different tensor properties, which is part of requirement \textbf{R4}.

To convey the available information in full detail, and in a way that enables the investigation of correlations, our system employs a scatter plot matrix, whose six dimensions reflect the IGRT
embedding that was introduced in Section~\ref{sec:igrt-projections}, so that they remain interpretable. As shown in Figure~\ref{fig:splom}, 15~scatter plots
visualize the correlations between any pair of IGRT axes. Each subject
is shown as a point, and colored according to its group (here:
patients red, healthy controls blue). For example, the top left
scatter plot in Figure~\ref{fig:splom} shows a strong negative
correlation between changes in Frobenius norm and Fractional
Anisotropy. When the mouse pointer hovers over a plot, the exact
Pearson correlation coefficient is displayed as a tooltip.

On the diagonal of the scatter plot matrix, we use boxplots to show
the distributions along the respective axis in the two
groups. This immediately reveals which axes exhibit a strong group
difference (in Figure~\ref{fig:splom}, norm and FA), and for which
others the groups overlap. Here,
tooltips contain the results of a univariate test, answering the
question if the respective dimension alone would have been sufficient for statistical significance.

The scatter plot matrix can be used in two ways: The first is simply
to browse the data, by clicking on any voxel in a spatial view. The
scatter plot matrix then displays the local tensor data from all
subjects. The second is for a post-hoc analysis of spatial clusters
which were previously highlighted by a hypothesis test. In this case,
the scatter plot matrix reflects the data from all subjects after
averging over the respective cluster. Special care has to be taken
when averaging values from axes 4--6, which correspond to the rotation
tangents. Since they are derived from eigenvectors, which lack an
intrinsic orientation, their signs could flip randomly from one voxel
to the next. Therefore, we average the absolute values in these
cases. The results still reflect the amount, but no longer the
direction of rotations.

\subsection{Small Multiple Glyph View}
\label{sec:glyph-view}

\begin{figure}
 \centering
 \includegraphics[width=\linewidth]{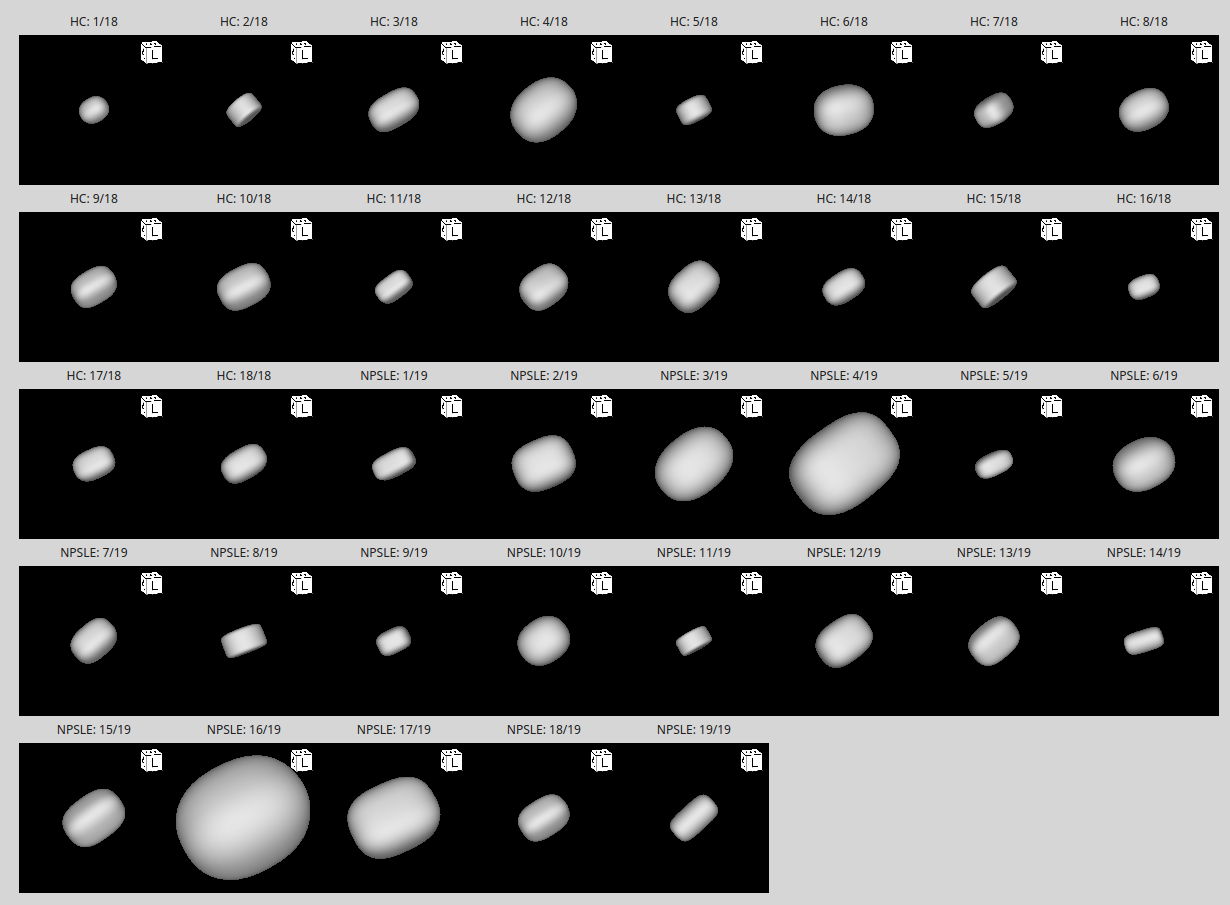}
 \caption{A small multiple glyph view shows the local tensor for each individual subject and allows for the easy assessment of residual misalignment. In this example, some tensors have much higher norm than most others, and immediately stand out as outliers.}
 \label{fig:glyph-view}
\end{figure}

Even though our clinical co-authors were quickly convinced that the scatter plot matrix excels at showing correlations, they pointed out that it remains rather abstract. Especially for requirement \textbf{R5,} i.e., the assessment of residual misalignments or cases of failed image registration, they still preferred to see a more traditional tensor glyph visualization.

Our system provides a small multiple glyph view, shown in Figure~\ref{fig:glyph-view}, which contains one tensor glyph per subject. In contrast to recently proposed glyphs that summarize groups of tensor fields \cite{zhang:2017ensemble}, this immediately reveals the identity of potential outliers. This is important so that the analyst can re-attempt a failed registration with modified parameters or, if no reasonable registration can be achieved, exclude the affected subject from the study.

Each subwindow is labeled with the respective subject ID, and all camera parameters are coupled, so that glyphs can easily be compared from different viewing directions. Figure~\ref{fig:glyph-view} shows a voxel near the boundary between white matter and the corticospinal fluid. The considerable variation in tensor norm and amount of anisotropy that is evident from the glyphs is a telltale sign of an imperfect registration. In this case, the glyph view led the analyst to disregard a statistical difference that had been detected in this voxel.

\subsection{Comparative View}
\label{sec:comparative-views}

\begin{figure}
 \centering
 \includegraphics[width=0.64\linewidth]{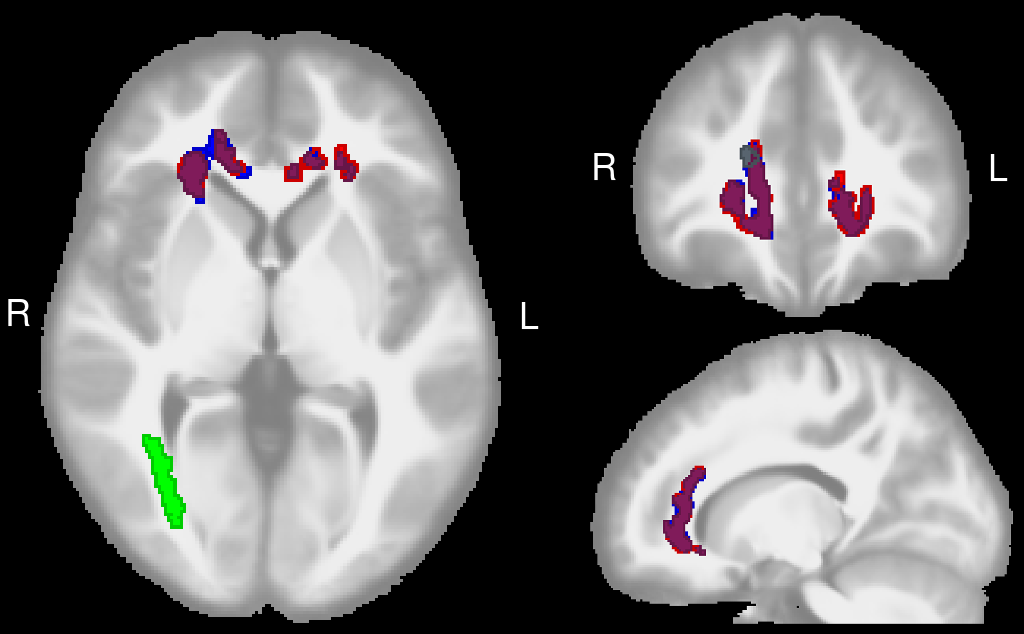}
 \hskip0.02\linewidth
 \raisebox{1ex}{\includegraphics[width=0.33\linewidth]{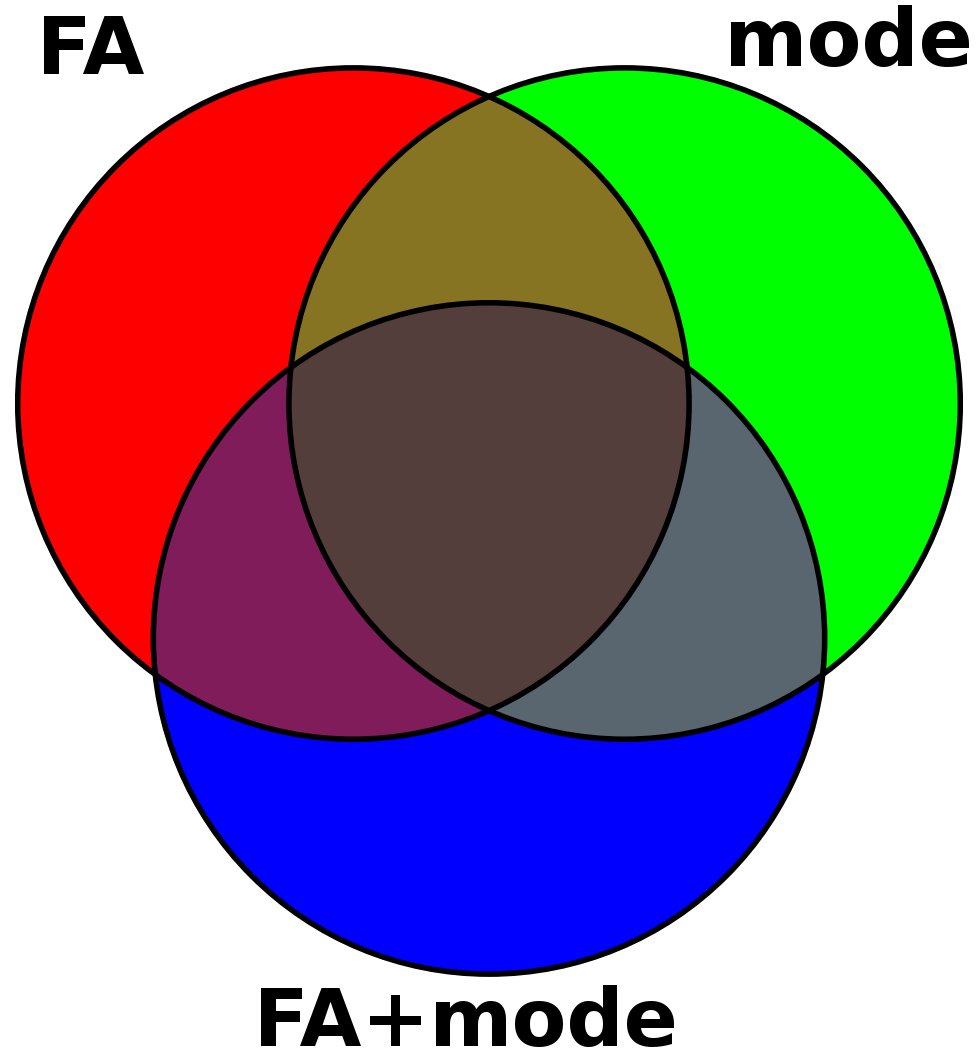}}
 \caption{A multivariate test FA+mode is compared to the two respective univariate tests by mapping the results of each to a different color, mixing them in a perceptual color space, and delineating them with contours. A Venn diagram serves as a color legend for regions of overlap.}
 \label{fig:overlay}
\end{figure}

To address requirement \textbf{R6,} our framework supports the comparison of results from different tests, thresholds, or levels of data
smoothing. After running a test, optionally with cluster enhancement,
one can save the resulting binary mask to a file. Up to three different test results can then be loaded into an overlay map, shown in Figure~\ref{fig:overlay}.

It is a difficult task to visually encode the overlap of different classes in a manner that is intuitive and easy to interpret. After trying out several alternatives, we chose an encoding that has been proposed in a very different application context, namely, in the ``splatterplot'' approach to overcoming overdraw in scatter plots \cite{Gleicher:2013}. This encoding combines specific rules for color blending and modulation with contouring.

Colors that encode regions of overlap are obtained in two steps: First, the colors that represent the overlapping classes are averaged in the CIE Lab color space to obtain a color that has approximately the same perceptual distance to all involved classes. Second, in order to more clearly indicate the presence of multiple classes, lightness and chroma of the mixed color are attenuated in the LCH color space, with an increasing effect for an increasing number of overlapping classes. The exact equations can be found in \cite{Gleicher:2013}, and the result can be observed in Figure~\ref{fig:overlay}. To further facilitate interpretation of the overlay map, our user interface displays a Venn diagram as a color legend.

Despite the advanced color blending and additional visual cues from contours, which we draw in slightly darker shades of the respective color, we came across cases in which the underlying maps were so complex that trying to understand the exact relationship of all three from a single image remained challenging. For these cases, our interface allows the user to temporarily hide some of the classes in order to build a better understanding in an interactive and iterative manner.

The exact number of voxels included in each mask, and the number of
voxels in the overlapping regions, are displayed as a tooltip of the
Venn diagram. We also update the table view to reflect the
connected components of the regions where the results of all three
tests overlap. Specific use cases, and interpretation of
Figure~\ref{fig:overlay}, are discussed in
Sections~\ref{sec:multivariate-tests} and~\ref{sec:data-smoothing}.

%%% Local Variables:
%%% mode: latex
%%% TeX-master: "../submission"
%%% End:

\section{Results And User Feedback}
\label{sec:results}

We designed, implemented, iteratively refined and extended the presented system over a two-year period during which we accounted for feedback from our clinical co-authors. Feedback from a case study is reported in Section~\ref{sec:case-study}. In the remainder of this section, we report specific results that have been achieved with our system. We focused on 56 diffusion tensor fields from a clinical study of systemic lupus erythematosus (SLE), for which we had previously reported results from a standard analysis in a medical journal paper that also provides detailed
information about the data acquisition, clinical parameters, and
criteria for inclusion \cite{schmidt2014diminished}. Among
the 56 subjects, 38 suffered from the disease, 19 of them with
neuropsychiatric symptoms (NPSLE), 19 without (non-NPSLE). The
remaining 18 subjects were in a healthy control group.

\subsection{Feedback From Case Study}
\label{sec:case-study}

We had a two-hour in-person meeting in which we provided a live demo of the system to one of our collaborators (TSW) and went through several examples together, discussing both the design of our software and the clinical implications of our findings.

Our collaborator particularly liked the fact that our system tightly integrates different techniques for statistical and quantitative data analysis with visualization, which is not achieved in the standard software packages available to him, and which encourages data exploration and hypothesis generation. He considered this helpful for DTI analysis, particularly highlighting the ability to easily assess correlations between different tensor properties, the option to directly seed tractography from regions in which significant differences were found, and the fact that the system displays quantitative results on demand, e.g., $t$ scores and $p$ values as tooltips in the SPLOM. He also emphasized that he would like to see similar systems for other modalities, such as functional MRI or Voxel Based Morphometry.

In preparation for the meeting, we reorganized our original user interface by placing controls that we considered to be less crucial to the main workflow into an optional ``advanced'' view, as indicated in Figure~\ref{fig:teaser}. Our collaborator found that, despite this effort, a certain level of complexity remained, but he thought it was commensurate with the complexity of the analysis task and told us that a certain amount of training is also needed to become familiar with the standard software packages in the field. He found the structure of the interface logical, and he asked for a few changes that we made in the final version, e.g., by adding the glyph view described in Section~\ref{sec:glyph-view}.

\subsection{Hypothesis Generation on Clinical Data}
\label{sec:hypothesis-generation}

\begin{figure}
  \centering
  \begin{minipage}[t]{0.9\linewidth}
  \includegraphics[width=\linewidth]{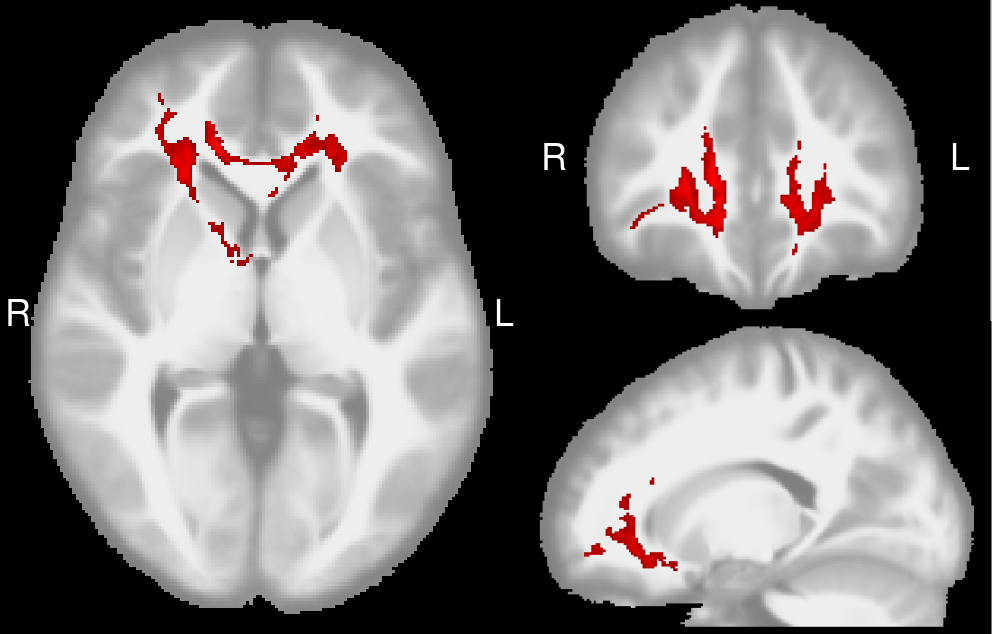}
  \centering\footnotesize (a) Fractional Anisotropy
  \end{minipage}%\hspace{1ex}
  \vskip0.5\baselineskip
  \begin{minipage}[t]{0.9\linewidth}
  \includegraphics[width=\linewidth]{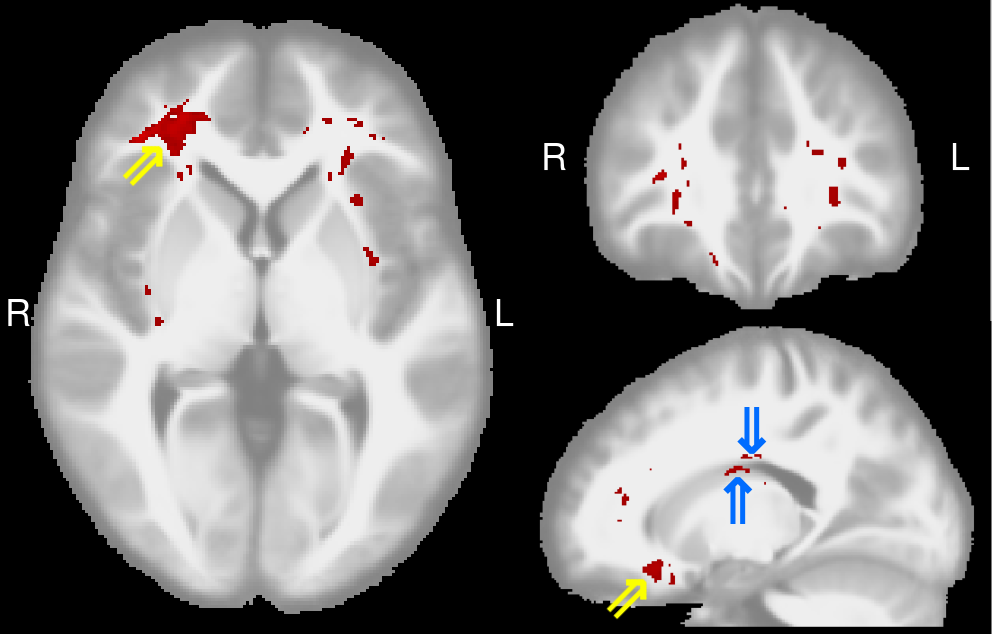}
  \centering  \footnotesize (b) Frobenius Norm
  \end{minipage}
  \vskip0.5\baselineskip
  \begin{minipage}[t]{0.9\linewidth}
  \includegraphics[width=\linewidth]{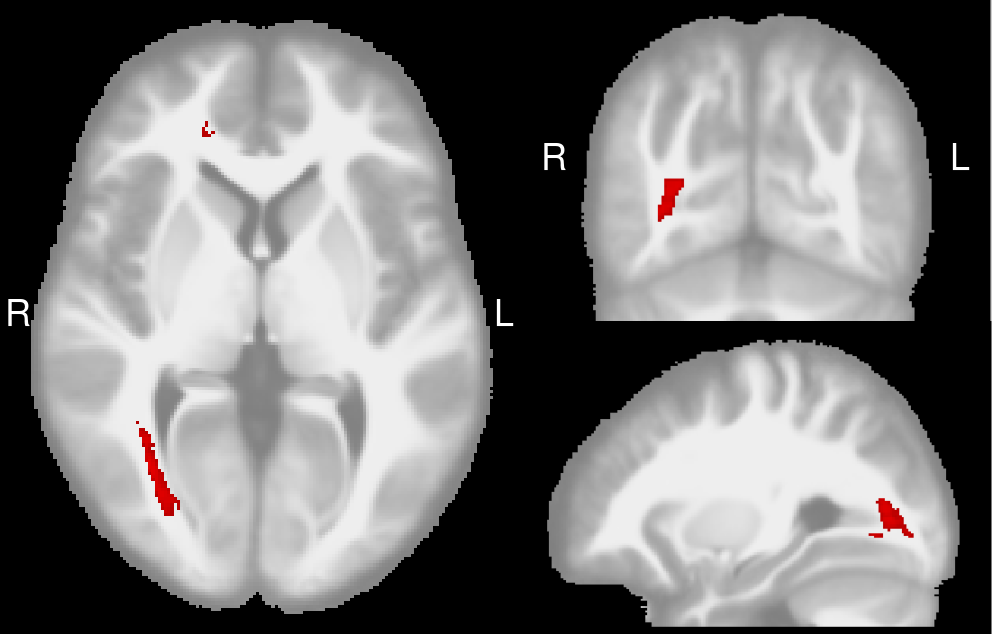}
  \centering\footnotesize (c) Tensor Mode
  \end{minipage}\hspace{1ex}
  \vskip0.5\baselineskip
  \begin{minipage}[t]{0.9\linewidth}
  \includegraphics[width=\linewidth]{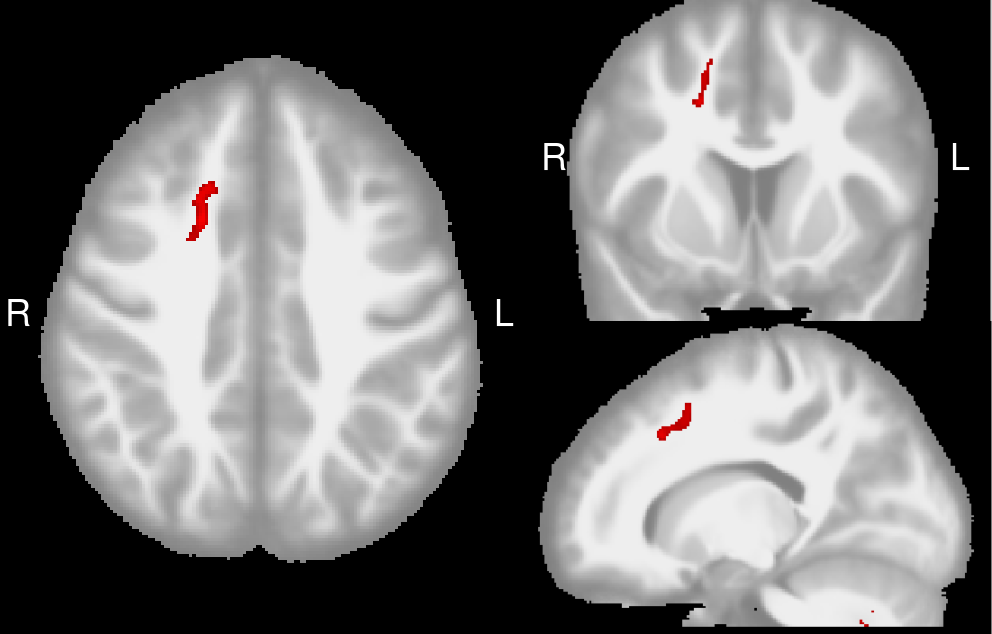}
  \centering  \footnotesize (d) Rotation around main axis
  \end{minipage}
  \caption{In data from a study on systemic lupus erythematosus (SLE),
    results for Fractional Anisotropy differences between healthy
    controls and NPSLE patients agree with previously published
    results, which validates our approach. Additional hypothesis tests
    can be constructed using our tool, and lead to interesting new
    observations.}
  \label{fig:hc-npsle}
\end{figure}

As it was explained in Section~\ref{sec:igrt-projections}, our
framework tests \emph{changes in tensor values} associated with certain
tensor invariants or rotations. This makes a univariate test of
Fractional Anisotropy (FA) in our tool similar, but not equivalent
to traditional hypothesis tests, which are based on \emph{changes in (pre-computed) FA.} Therefore, comparing the results to those from a traditional
analysis can serve as a valuable initial
validation. Figure~\ref{fig:hc-npsle}~(a) shows the result from
comparing healthy controls to NPSLE patients in our framework, based
on FA-related changes and threshold-free cluster enhancement. The
detected regions are remarkably similar to those from our previous
analysis of the same data with standard methods
\cite{schmidt2014diminished}. Seeding tractography in them also
reveals most of the fibers that have been reported in a previous
tractography-based study \cite{Shastri2016} and have been ranked
highly in terms of their predictiveness \cite{Khatami:2017}, including
the genu of the corpus callosum, left and right inferior
fronto-occipital fasciculus, and left uncinate fasciculus (see also Figure~\ref{fig:fibers}).

The remaining subfigures of Figure~\ref{fig:hc-npsle} show the results
from other univariate hypothesis tests that are enabled by our
framework, and for which no similar results have been reported
previously. Subfigure~(b) shows regions in which the Frobenius norm of
the diffusion tensors is increased in the patients, indicating
increased overall diffusivity. The largest cluster (yellow arrows) is immediately adjacent to, and partly overlapping with, the previously detected changes in FA. Only a small part of the second largest cluster is visible in Figure~\ref{fig:hc-npsle}~(b) (blue arrows). It is directly
adjacent to a ventricle, and further inspection with the glyph view indicates that this second cluster might be caused by a difference in
ventricle sizes between the groups, which might have been compensated only
partially by the registration \cite{smith2006tract}.

Subfigure~(c) shows a cluster in which tensor mode is decreased
in the patients, which means that diffusion is restricted more to a
plane than a line, and which might indicate a larger degree of fiber
spread in the patients. A tractography result seeded in this cluster indicates that the difference extends over parts of the forceps major and inferior fronto-occipital fasciculus.

The only result we observed when testing for differences in
orientation was a cluster in the corona radiata of the right
hemisphere, where tensors are systematically twisted (rotated around
their principal eigenvector) in the patient group compared to the
controls. It is unclear whether this is an artifact of the specific
sample or the registration algorithm, or if it corresponds to a true
difference between the populations, especially given that the cluster
is small.

In summary, our framework was successfully used to produce new
hypotheses that should be investigated further in future clinical
studies.

\subsection{Exploring Results of Multivariate Tests}
\label{sec:multivariate-tests}

All experiments in the previous subsection used univariate hypothesis tests. To illustrate use of the scatter plot matrix to investigate the results from a multivariate test, we looked for any differences in tensor shape (i.e., in the subspace spanned by norm, FA, and mode) between the healthy controls and the non-NPSLE patients, again using threshold-free cluster enhancement (TFCE).

As it can be seen in Figure~\ref{fig:splom}, this results in several medium-sized clusters. Three of them, shown in red, green, and purple in the inset slice views, exhibit significantly reduced FA ($t=-4.61$ in the largest, red cluster), and increased Frobenius norm ($t=4.38$) in the patient population, with a strong correlation between both measures ($\rho=-0.91$). Again, these might be due to differences in ventricle size. In the blue cluster, FA is strongly decreased in non-NPSLE patients ($t=-4.14$), but norm is less affected ($t=1.63$), and the correlation is much weaker ($\rho=-0.23$). Decreased FA in the same region was previously reported based on standard methods \cite{schmidt2014diminished}. In addition, our framework allows us to observe that this difference goes along with a reduction in mode ($t=-2.51$). 

One question that the comparative view from Section~\ref{sec:comparative-views} allows us to answer is how the results from two univariate tests, e.g., ones that compare NPSLE patients to healthy controls based on FA or tensor mode, differ from a single multivariate test that combines both. Figure~\ref{fig:overlay} shows the corresponding overlay.

As it was mentioned in Section~\ref{sec:tfce-enhancement}, our software does not convert the results from TFCE to $p$ values, since this would require time consuming permutation testing. In order to still compare univariate and multivariate tests in a meaningful manner, Figure~\ref{fig:overlay} was created using the Hotelling test without TFCE. In this case, the null distribution has a parametric form \cite{Srivastava:2002}, which allows us to compute $p$ values within interactive runtimes. To compensate for the fact that testing FA and mode separately amounts to performing twice as many tests as a single combined test, we made the thresholds for the univariate tests twice as restrictive ($p_\text{uncorr}<0.005$ vs.\ $p_\text{uncorr}<0.01$).

In this example, the comparative view revealed that the volume highlighted by the multivariate test ($7.8\,\mathrm{cm}^3$) mostly agrees with the union of the two univariate tests ($4.3\,\mathrm{cm}^3$ for FA and $4.0\,\mathrm{cm}^3$ for mode), with almost no overlap between FA and mode. This leads us to the hypothesis that testing mode in future studies of SLE will result in new findings that complement the ones from FA.

\subsection{Visualizing the Effect of Data Smoothing}
\label{sec:data-smoothing}

There has been some controversy about whether to smooth diffusion tensor data in preparation for statistical analysis. On one hand, spatial smoothing helps to compensate for anatomical misalignment that may remain after registration, it smoothes out noise that might otherwise lead to false positive detections, it makes the data distribution more Gaussian and, when the bandwidth is matched to the spatial scale of regions of difference, it can boost statistical power. On the other hand, the ideal spatial scale is usually not known in advance, and there is no principled approach for selecting the bandwidth parameter, which might have a substantial effect on the results \cite{JONES2005}. Moreover, it has been argued that, since smoothing amounts to artificially decreasing the image resolution, it is counter-productive for diffusion MRI in particular, which specifically measures diffusion in order to overcome the limits of image resolution and to infer tissue parameters at a microscale \cite{smith2006tract}.

For these reasons, Tract-Based Spatial Statistics (TBSS), a widely used method for statistical hypothesis testing of diffusion tensor fields, avoids spatial smoothing and instead corrects for residual misalignment by projecting FA values onto a so-called white matter skeleton, a medial surface representation of the major fiber tracts \cite{smith2006tract}. Unfortunately, current version of TBSS cannot deal with the full tensor information. Therefore, recent studies that have analyzed the full tensor have still employed smoothing \cite{bouchon2016most}.

\begin{figure}
  \centering
  \begin{minipage}[t]{0.9\linewidth}
  \includegraphics[width=\linewidth]{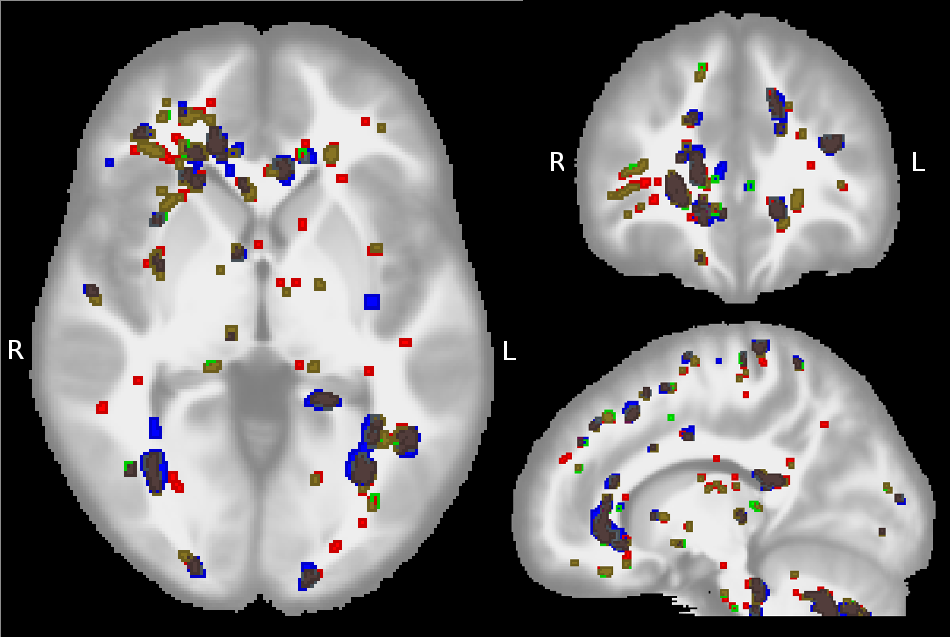}
  \centering\sffamily\footnotesize Results without TFCE
  \end{minipage}
  \vskip0.5\baselineskip
  \begin{minipage}[t]{0.9\linewidth}
  \includegraphics[width=\linewidth]{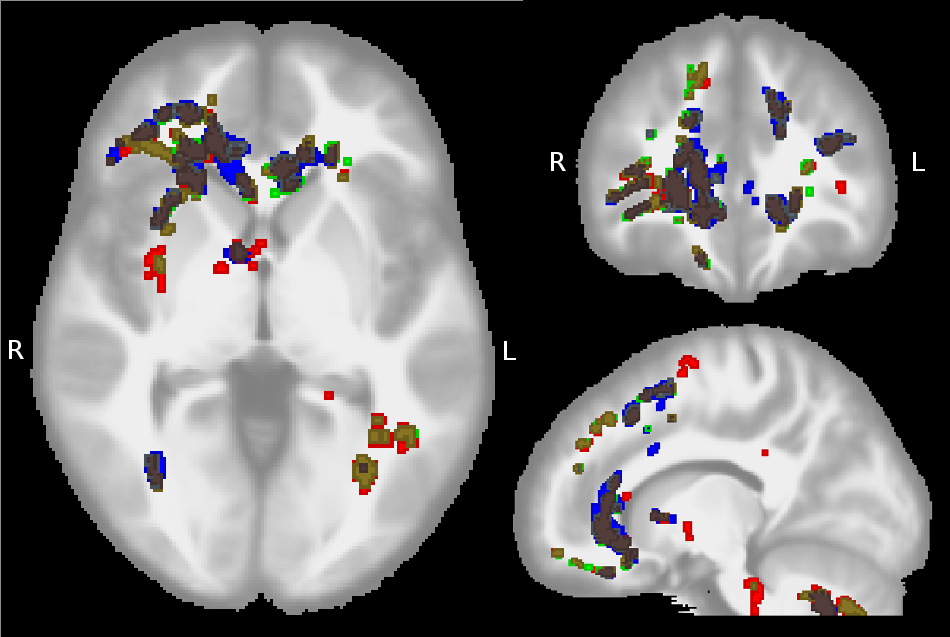}
  \centering\sffamily\footnotesize Results with TFCE
  \end{minipage}
 \caption{Our decision to use threshold-free cluster enhancement (TFCE) is based on comparing results with (bottom row) and without it (top row). The color scheme is as in Fig.~\ref{fig:overlay}, with red representing results without any smoothing, green with $\sigma=0.7$, blue with $\sigma=1.7$. TFCE reduces the effect, and also the necessity, of pre-smoothing.}
 \label{fig:smoothing}
\end{figure}

We have used the comparative view from Section~\ref{sec:comparative-views} to explore the effects of data smoothing, as shown in Figure~\ref{fig:smoothing}. In this experiment, we compared the healthy and NPSLE populations with respect to the full tensor information. As in \cite{bouchon2016most}, we compare the spatial regions that result from different processing pipelines by showing the $5\%$ ``most significant'' voxels, rather than setting some \emph{a priori} threshold.

The top row of Figure~\ref{fig:smoothing} compares results on the original data (red) to data that has been smoothed with bandwidth $\sigma=0.7$ voxels (green) and $\sigma=1.7$ voxels (blue), respectively. Many isolated voxels and small regions are shown in red, indicating that they are removed by smoothing, while many of the larger ones grow, as indicated by blue halos around them. Optionally, the analyst can focus on the dark brown regions, indicating differences that are considered significant irrespective of the amount of smoothing.

The bottom row shows results from the same experiment, but additionally uses threshold-free cluster enhancement (TFCE). Even without any smoothing, TFCE eliminates many of the small clusters, and leads to larger connected ones. It also reduces the overall effect of smoothing, leading to a $10\%$ increase in the number of consensus voxels. Based on these results, we decided to omit smoothing, but use TFCE in most experiments reported above.

%%% Local Variables:
%%% mode: latex
%%% TeX-master: "../submission"
%%% End:

\section{Conclusion}
\label{sec:conclusion}

Diffusion tensor MRI is widely used for studying how different brain diseases affect white matter microstructure. The multivariate nature of the resulting tensor fields makes it a challenge to comprehensively explore group differences. In this paper, we have argued to approach this task using visual analytics, based on the observation that judging which of the differences are likely to be related to the respective disease requires combining the analyst's prior anatomical knowledge with complex statistical computation.

We presented a tool that builds on the formalism of statistical hypothesis testing and provides a suitable visual interface to specify different null hypotheses and to explore the spatial extent of the resulting significant regions, based on the assumption that larger connected clusters are more likely to be relevant to the analyst. By linking them to three-dimensional visualizations such as fiber tractography, it becomes easy to identify the affected bundles. We also provide support for \emph{a posteriori} investigation of the results from multivariate tests, allowing us to identify which specific attributes of the tensor fields exhibit the strongest differences, and how they are correlated. Finally, overlay views make it easy to compare the result from different tests, thresholds, or levels of smoothing. Throughout this pipeline, we closely integrate visual with quantitative analysis.

In terms of the specific results shown on data from a clinical study on systemic lupus erythematosus, a logical next step will be to verify, on data from an independent cohort, the new hypotheses that we arrived at by using our tool, in particular the hypotheses that the disease also goes along with changes in overall diffusivity, as well as with changes in tensor mode in regions that are disjoint from the ones in which Fractional Anisotropy changes.

We note that, even though the vast majority of hypothesis testing in clinical studies that use diffusion MRI is still based on the diffusion tensor model, this model has long been known to be insufficient for modeling multiple fiber directions within the same voxel, and high-angular resolution diffusion imaging is now considered to be a state-of-the-art alternative for fiber tractography \cite{Jeurissen:NMR2017}. More recently, so-called multi-shell models, such as diffusional kurtosis \cite{Jensen:2010} or NODDI \cite{Zhang:2012}, which provide even more detailed information, have started to enter clinical research, and only very few visualization techniques are currently available for them \cite{Bista:2014}. This is a very natural direction for future extensions of our framework.

%In the more general area of supporting image-based research in medicine and biology, we observe that the big data challenges that arise from high-throughput imaging and that require novel visual analytics technology for data analysis affect a wide range of research questions, from connectomics on a cellular \cite{Lichtman2014} to a whole-brain level \cite{Sotiropoulos:2013}
%and from fundamental questions of biochemistry \cite{Zadeh2016} to epidemiological studies \cite{Klemm2014}. We expect that this will continue to stimulate much visualization research in the near future.

%%% Local Variables:
%%% mode: latex
%%% TeX-master: "../submission"
%%% End:

%\bibliographystyle{eg-alpha}
%\bibliographystyle{eg-alpha-doi}

\bibliographystyle{IEEEtran}
\bibliography{submission}

% Can use something like this to put references on a page
% by themselves when using endfloat and the captionsoff option.
\ifCLASSOPTIONcaptionsoff
  \newpage
\fi

\end{document}